\documentclass[conference]{IEEEtran}
\IEEEoverridecommandlockouts

\usepackage{listings}
\usepackage{xcolor}
\usepackage{verbatim}
\usepackage{multirow}
\usepackage{graphicx}
\usepackage{url}
\usepackage{wasysym}
\usepackage{textcomp}
\usepackage{subcaption}
\usepackage{amssymb}
\usepackage{cite}
\usepackage{balance}
\usepackage[top=0.7in, bottom=1.09in, left=0.68in, right=0.68in]{geometry}

\makeatletter
\def\hlinew#1{%
  \noalign{\ifnum0=`}\fi\hrule \@height #1 \futurelet
   \reserved@a\@xhline}

\newcommand{\tabincell}[2]{\begin{tabular}{@{}#1@{}}#2\end{tabular}}
\renewcommand{\baselinestretch}{0.99}
\begin{document}

%
\title{Under-Optimized Smart Contracts\\Devour Your Money}


%


\author{
\IEEEauthorblockN{
Ting Chen\IEEEauthorrefmark{1}\IEEEauthorrefmark{2}, 
Xiaoqi Li\IEEEauthorrefmark{2}, 
Xiapu Luo\IEEEauthorrefmark{2}\thanks{\IEEEauthorrefmark{3} The corresponding author.}\IEEEauthorrefmark{3},
Xiaosong Zhang\IEEEauthorrefmark{1}
}
\IEEEauthorblockA{
\IEEEauthorrefmark{1}Center for Cybersecurity, University of Electronic Science and Technology of China, China\\
\IEEEauthorrefmark{2}Department of Computing, The Hong Kong Polytechnic University, China\\
Email: brokendragon@uestc.edu.cn, csxqli@gmail.com, csxluo@comp.polyu.edu.hk, johnsonzxs@uestc.edu.cn}
}

\maketitle
\begin{abstract}\label{abs}Smart contracts are full-fledged programs that run on blockchains (e.g., Ethereum, one of the most popular blockchains). In Ethereum, gas (in Ether, a cryptographic currency like Bitcoin) is the execution fee compensating the computing resources of miners for running smart contracts. However, we find that under-optimized smart contracts cost more gas than necessary, and therefore the creators or users will be overcharged. In this work, we conduct the \emph{first} investigation on Solidity, the recommended compiler, and reveal that it fails to optimize gas-costly programming patterns. In particular, we identify 7 gas-costly patterns and group them to 2 categories. Then, we propose and develop \emph{GASPER}, a new tool for automatically locating gas-costly patterns by analyzing smart contracts' bytecodes. The preliminary results on discovering 3 representative patterns from 4,240 real smart contracts show that 93.5\%, 90.1\% and 80\% contracts suffer from these 3 patterns, respectively.
\end{abstract}

\section{Introduction}
\label{intro}
The success of Bitcoin, a decentralised cryptographic currency that reached a capitalisation of 10 billions of dollars since its launch in 2009~\cite{bitcoin_money}, has attracted lots of attentions from both industry and academia to investigate the underlying technology of cryptocurrencies, the blockchain. One prominent  application on blockchains is to execute smart contracts, which can be considered as full-fledged programs running on blockchains from the perspective of software engineering.

Ethereum is one of the most popular blockchains where more than 10 million transactions had occurred~\cite{trans} . The term ``blockchain'' and ``smart contract'' refer to the Ethereum blockchain and its smart contracts, respectively, below without special declaration. A smart contract can be developed in Solidity (the recommended language), Serpent, or LLL. No matter which programming language is used, the source of a smart contract will be complied into  bytecodes that can be executed in the Ethereum Virtual Machine (EVM for short).

Smart contracts run on the machines of miners, who can earn Ethers (i.e., the cryptographic currency circulated in Ethereum)  by contributing their computing resources. The creators and users of smart contracts will be charged certain amount of gas for purchasing the computing resources from miners. The charge of a transaction equals to the multiplication of the gas consumed by executing the transaction and the price of gas (Ether per unit).  Moreover, when deploying contracts, the creators will also be charged of gas, the amount of which are related to the size of smart contracts in bytecodes.


We find that under-optimized smart contracts cost more gas than necessary, and therefore the creators or users will be overcharged. To save money, developers had better follow gas-efficient programming patterns. Unfortunately, there is not such a guideline yet, and it is difficult for developers to identify gas-costly bytecode and replace them with gas-efficient ones, because it requires deep understanding of EVM's instructions, the gas consumption for different operations, the data locations accessed by operations, the amount of data read or written etc. Hence, a compiler that can optimize the bytecode for minimizing gas consumption is highly desired.

In this paper, we conduct the \emph{first} investigation on Solidity, the recommended compiler for Ethereum, and reveal that it fails to optimize gas-costly programming patterns. More precisely, we identify 7 gas-costly patterns and divide them into 2 categories: useless-code related patterns, and loop-related patterns. Furthermore, we propose and develop \textsc{Gasper} (short for GAS-costly Patterns checkER), a new tool for discovering gas-costly patterns in bytecode automatically. \textsc{Gasper} leverages symbolic execution and it currently can locate 3 representative patterns, which cover the two categories. By applying \textsc{Gasper} to analyze all deployed smart contracts until Nov. 5th, 2016, we find that 93.5\%, 90.1\% and 80\% smart contracts suffer from these 3 patterns, respectively. It is worth noting that although the list of our patterns is by no means of complete, this research sheds light on this important issue and hopefully stirs more research on it.

Overall, we make the following contributions:
\begin{itemize}
	\item To our best knowledge, this is the \emph{first} investigation revealing that lots of smart contracts, generated by the recommended compiler,  contain gas-costly bytecodes, which can be replaced with gas-efficient bytecodes to save money.
	\item We propose and develop \textsc{Gasper}, a new tool based on symbolic execution for automatically discovering gas-costly patterns in bytecode. The current version covers 3 representative patterns in 2 categories, and is being extended to support more patterns.
	\item We apply \textsc{Gasper}  to all deployed smart contracts until Nov. 5th, 2016, and find that 93.5\%, 90.1\% and 80\% smart contracts suffer from these 3 patterns, respectively.
\end{itemize}

\section{Background}
\label{gas}
Gas is used for purchasing computing resources from miners since smart contracts run on miners' machines. Gas can be considered as money with equivalent value. For example, the average gas price on Nov. 11, 2016 is 0.000000024334480804 Ether~\cite{gasprice}, which is roughly equal to $2.5\times 10^{-7}$ US dollars~\cite{exchange}. Note that the gas price and the exchange rate of Ether to US dollar are determined by the market and keep changing. 

Deploying and executing smart contracts cost money. For instance, an addition operation that sums up the top two items of the stack takes 3 units of gas, about $7.5\times 10^{-7}$ US dollars. One may argue that the cost for an addition is so low that we do not need to optimize it. However, it is worth noting that real smart contracts consist of lots of operations and some operations consume much more gas than the addition operation, as shown in Table \ref{t_gas}. Moreover, smart contracts usually provide public methods that can be called unlimited times by various clients and contracts. Hence, an optimized smart contract can save obvious gas (i.e., money) than its un-optimized counterpart due to the scale effect.

\begin{table}[ht!]
	\centering
	\vspace*{-1ex}
	\scriptsize
	\caption{Gas cost of different operations, a complete list can be found in Ethereum's yellow paper~\cite{gas}}
	\vspace{-1ex}
	\label{t_gas}
	\begin{tabular}{|c|c|c|}
		\hline
		\textbf{Operation}        & \textbf{Gas}                                                   & \textbf{Description}                                                                           \\ \hline
		\textsc{ADD/SUB}          & 3                                                     & \multirow{3}{*}{Arithmetic operation}                                                           \\ \cline{1-2}
		\textsc{MUL/DIV}          & 5                                                     &                                                                                       \\ \cline{1-2}
		\textsc{ADDMOD/MULMOD}    & 8                                                     &                                                                                       \\ \hline
		\textsc{AND/OR/XOR}       & 3                                                     & Bitwise logic operation                                                                             \\ \hline
		\textsc{LT/GT/SLT/SGT/EQ} & 3                                                     & Comparison operation                                                                            \\ \hline
		\textsc{POP}              & 2                                                     & \multirow{2}{*}{Stack operation}                                                                \\ \cline{1-2}
		\textsc{PUSH/DUP/SWAP}    & 3                                                     &                                                                                       \\ \hline
		\textsc{MLOAD/MSTORE}     & 3                                                     & Memory operation                                                                                \\ \hline
		\textsc{JUMP}             & 8                                                     & Unconditional jump                          \\ \hline
		\textsc{JUMPI}            & 10                                                    & Conditional jump                            \\ \hline
		\textsc{SLOAD}            & 200                                                   & \multirow{2}{*}{Storage operation}                                                              \\ \cline{1-2}
		\textsc{SSTORE}           &\tabincell{c}{5,000/\\20,000}&                                                                                       \\ \hline
		\textsc{BALANCE}          & 400                                                   & Get balance of an account                  \\ \hline
		\textsc{CREATE}           & 32,000                                                 & Create a new account using CREATE             \\ \hline
		\textsc{CALL}             & 25,000                                                 & Create a new account using CALL            \\  \hline
	\end{tabular}
	\vspace{-2ex}
\end{table}

Stack operations (e.g., \texttt{POP}, \texttt{PUSH}), arithmetic operations (e.g., \texttt{ADD}, \texttt{SUB}), bitwise operations (e.g., \texttt{OR}, \texttt{XOR}), and comparison operations (e.g., \texttt{LT}/\texttt{GT}) are cheap because being a stack-based virtual machine, EVM favors such stack-related operations. Loading a word (i.e., 256 bits) from the memory (e.g., \texttt{MLOAD}) or saving a word to the memory (e.g., \texttt{MSTORE}) are also cheap. The term ``memory'' referred in Ethereum stands for a special memory area, of which a contract obtains a freshly cleared instance for each message call. For example, the data attached in a message call is stored in memory. It is worth noting that the gas consumption will be multiplied if many words in memory are read or written. Moreover, memory can be expanded when accessing a previously untouched memory location. Every expanded word needs 3 units of gas.

Loading a word from the storage (i.e., \texttt{SLOAD}) or saving a word to the storage (i.e., \texttt{SSTORE}) are expensive. The term ``storage'' referred in Ethereum is a persistent memory area where any changes to the storage by one call of a contract can be observed by subsequent calls of that contract. A \texttt{SSTORE} operation costs 20,000 units of gas if the storage word is set to non-zero from zero; otherwise, it costs 5,000. It is worth noting that although the caller of a contract will be refunded 15,000 units of gas if a \texttt{SSTORE} operation sets a non-zero storage word to zero, the refund will not be committed until the transaction completes successfully.

EVM has a number of blockchain-specific operations which are very expensive, such as \texttt{BALANCE}, \texttt{CREATE} and \texttt{CALL}. Moreover, a conditional jump (i.e, \texttt{JUMPI}) is more expensive that an unconditional jump (i.e., \texttt{JUMP}). The gas consumption of each operation is susceptible to change due to the fast evolving of Ethereum. Roughly speaking, users are charged proportionally to the consumed computing resources.

\section{Gas-costly Programming Patterns}
\label{pattern}

We identify 7 gas-costly patterns, which can be classified into two categories: useless code related patterns and loop related patterns. The former introduces additional cost due to the increased size of bytecode during the deployment and the removable bytecode in runtime. The latter involves using expensive operations in the loop.  We have validated all these patterns using the latest Solidity (V 0.4.4) whose optimization is enabled. More precisely, we feed Solidity the gas-costly patterns in source code, and then check whether the gas-costly patterns are converted into gas-efficient ones in the generated bytecode.  The results show that \emph{none} of these patterns has been optimized by Solidity. For the ease of illustration, we present the patterns in source code rather than bytecode.
\vspace{-1ex}
\subsection{Category 1: Useless Code Related Patterns}

\begin{figure}[ht]
	\centering
	\vspace*{-2ex}
	\includegraphics[width=2.50in]{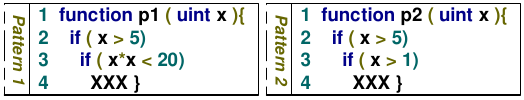}
	\vspace*{-1ex}
	\caption{Pattern 1: dead code, and Pattern 2: opaque predicate}
	\vspace{-2ex}
	\label{p12}
\end{figure}
1) Dead code. Fig.\ref{p12} (Pattern 1) gives an example of dead code where Line 4 will not be executed because the predicate ``x*x$<$20'' at Line 3 is evaluated to false under all circumstances.  Solidity does not remove Line 3 and 4 from the generated bytecode and hence wastes money.

2) Opaque predicate. The outcome of an opaque predicate is known to be true or false without execution. For example, the predicate ``x$>$1'' in Fig.\ref{p12} (Pattern 2) is an opaque predicate. 
Since the predicate at Line 3 is evaluated to true under all circumstances, it should be removed for saving gas. 
\vspace{-1ex}
\subsection{Category 2: Loop Related Patterns}

\begin{figure}[ht]
	\centering
	\vspace*{-2ex}
	\includegraphics[width=3.00in]{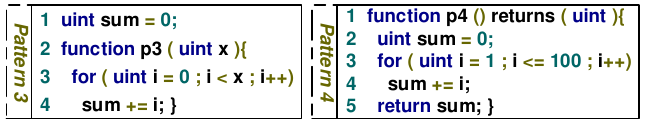}
	\vspace*{-1ex}
	\caption{Pattern 3: expensive operations in a loop, and Pattern 4: constant outcome of a loop}
	\vspace{-2ex}
	\label{p34}
\end{figure}

1) Expensive operations in a loop. The expensive operations in a loop are worth attention because they may execute multiple times in one invocation. Moving the expensive operations out of the loop can save gas. For example, in Fig.\ref{p34} (Pattern 3), since the variable \textit{sum} is stored in the storage, Line 4 involves a \texttt{SLOAD} for loading \textit{sum} to the stack and a \texttt{SSTORE} for saving the outcome of the \texttt{ADD} to the storage. Note that the storage-related operations are very expensive.

An advanced compiler should assign \textit{sum} to a local variable (e.g., \textit{tmp}) that resides in the stack, then add \textit{i} to \textit{tmp} inside the loop, and finally assign \textit{tmp} to \textit{sum} after the loop. Such optimization reduces the storage-related operations from 2\textit{x} to just 2, i.e., one \texttt{SLOAD} and one \texttt{SSTORE}. 

2) Constant outcome of a loop. In some cases, the outcome of a loop may be a constant that can be inferred in compilation. As shown in Fig.\ref{p34} (Pattern 4), the storage variable \textit{sum} in \textit{p}4 equals to 5050 after the loop. Hence, the body of \textit{p}4 should be simplified as ``return 5050;''.

\begin{figure}[ht]
	\centering
	\vspace*{-1ex}
	\includegraphics[width=3.00in]{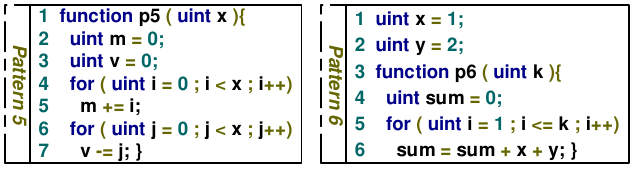}
	\vspace*{-1ex}
	\caption{Pattern 5: loop fusion, and Pattern 6: repeated computations in a loop}
	\vspace{-2ex}
	\label{p56}
\end{figure}

3) Loop fusion. It combines several loops into one if possible and thus reduces the size of bytecode. In particular, it can reduce the amount of operations, such as conditional jumps and comparison, etc., at the entry points of loops. The two loops shown in Fig.\ref{p56} (Pattern 5) can be combined into one loop, where both \emph{m} and \emph{v} get updated.

4) Repeated computations in a loop. In some cases, there may be expressions that produce the same outcome in each iteration of a loop. Hence, the gas can be saved by computing the outcome once and then reusing the value instead of recomputing it in subsequent iterations, especially, for the expressions involving expensive operands. For example, in Fig.\ref{p56} (Pattern 6), the gas consumption is very high due to the repeated computations. More precisely, the summation of two storage words (i.e., ``x+y'' at Line 6) is quite expensive because \textit{x} and \textit{y} should be loaded into the stack (i.e., \texttt{SLOAD}) before addition. To save gas,  this summation should be finished before the loop, and then the result is reused within the loop.



\begin{figure}[ht]
	\centering
	\vspace*{-1ex}
	\includegraphics[width=2.30in]{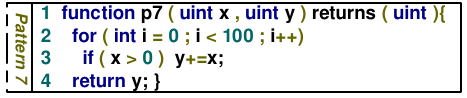}
	\vspace*{-1ex}
	\caption{Pattern 7: Comparison with unilateral outcome in a loop}
	\vspace{-2ex}
	\label{p9}
\end{figure}

5) Comparison with unilateral outcome in a loop. It means that a comparison is executed in each iteration of a loop but the result of the comparison is the same even if it cannot be determined in compilation (i.e., not an opaque predicate). For instance, in Fig.\ref{p9}, the comparison at Line 3 should be moved to the place before the loop.

\textit{Summary:} Adequate optimizations can reduce the cost of contract creators if the size of smart contracts can be reduced (e.g., eliminating dead code, removing unnecessary comparisons), and the cost of contract users if the computations of smart contracts can be reduced (e.g., moving expensive operations out of a loop). It is worth noting that the loop-related patterns will cost more gas with the increase of the loop count.

\section{Gasper}
\label{tool}
We propose and develop \textsc{Gasper} to automatically discover gas-costly programming patterns from the bytecode of smart contracts. \textsc{Gasper} handles bytecode directly without the need of source code, because only a few (728 until Nov. 29th, 2016) smart contracts open their sources. As an early research achievement, the current version of \textsc{Gasper} can find all patterns in category 1 and one representative pattern (i.e., expensive operations in a loop) in category 2. The detection of other patterns is in development.

\textsc{Gasper} conducts symbolic execution on bytecode to cover all reachable code blocks (a block is a straight-line code sequence with no branches in except to the entry and no branches out except at the exit). Given a smart contract, \textsc{Gasper} first disassembles its bytecode using \textit{disasm} provided by Ethereum.  Then, \textsc{Gasper} constructs the Control Flow Graph (CFG). It is worth noting that the CFG will be improved gradually during symbolic execution if new control flow transfers are found. Symbolic execution starts from the root node of the CFG, and traverses the CFG. If \textsc{Gasper} encounters a conditional jump, it checks which branches (i.e., true or false) are feasible by querying the Z3 solver~\cite{z3}. If both are feasible, \textsc{Gasper} selects one branch following  the depth-first search.



\subsection{Detection of Dead Code}
\label{detect_dead}
\textsc{Gasper} detects dead code through three steps. First, it logs the addresses of all executed blocks by symbolic execution. Then, it collects the addresses of all blocks by scanning the CFG. Finally, \textsc{Gasper} reports all blocks that are found in the CFG but not executed by symbolic execution as dead code. 


\subsection{Detection of Opaque Predicates}
\label{detect_opaque}
To detect opaque predicates, \textsc{Gasper} executes the smart contract symbolically, and records the executed branch (i.e., true or false) when a conditional jump is encountered. After that,  the conditional jump with one never-executed branch is regarded as an opaque predicate.

\subsection{Detection of Expensive Operations in a Loop}
\label{detect_expensive}
\textsc{Gasper} detects this pattern through two steps. First, \textsc{Gasper} looks for loops in the bytecode. Second, it searches loop bodies for expensive operations.  
More precisely, \textsc{Gasper} firstly searches for back edges in the CFG, which indicate the existence of loops, and then identifies the entry block and exit block for each loop. Afterwards, using Dijkstra algorithm, \textsc{Gasper} calculates the distances between each block with the entry block and exit block, respectively. The distance between two nodes is the least number of edges from one node to the other. A block is considered to be in a loop if it is closer to the exit block than to the entry block.  Currently, \textsc{Gasper} supports detecting 3 expensive operations, including \texttt{SLOAD}, \texttt{SSTORE} and \texttt{BALANCE}. More operations will be included in future work.


\section{Evaluation}
\label{evaluation}

We have implemented \textsc{Gasper}  based on \textsc{Oyente}\cite{oyente}, and evaluated it using all smart contracts deployed on Ethereum. More precisely, we scan all addresses in the blockchain because each deployed contract must be associated with an unique address. We find 566,907 addresses till November 5th, 2016, of which 539,617 addresses contain no bytecodes. Therefore, we download 27,290 contracts' bytecodes in total. Moreover, we find that many contracts are exactly the same (i.e., their bytecodes are identical). After eliminating identical contracts, 4,669 contracts are left. During experiments, 429 (less than 10\%) contracts cannot be examined because \textsc{Oyente} crashes due to its internal errors(e.g., \emph{Unknown Instructiondelegatecall}, \emph{Stack Underflow}, \emph{Unknown  Instructionextcodesize}) or \textsc{Oyente} runs out of time. Eventually, 4,240 contracts are successfully inspected.

\begin{figure}[ht]
	\centering
	\vspace*{-3ex}
	\includegraphics[width=2in]{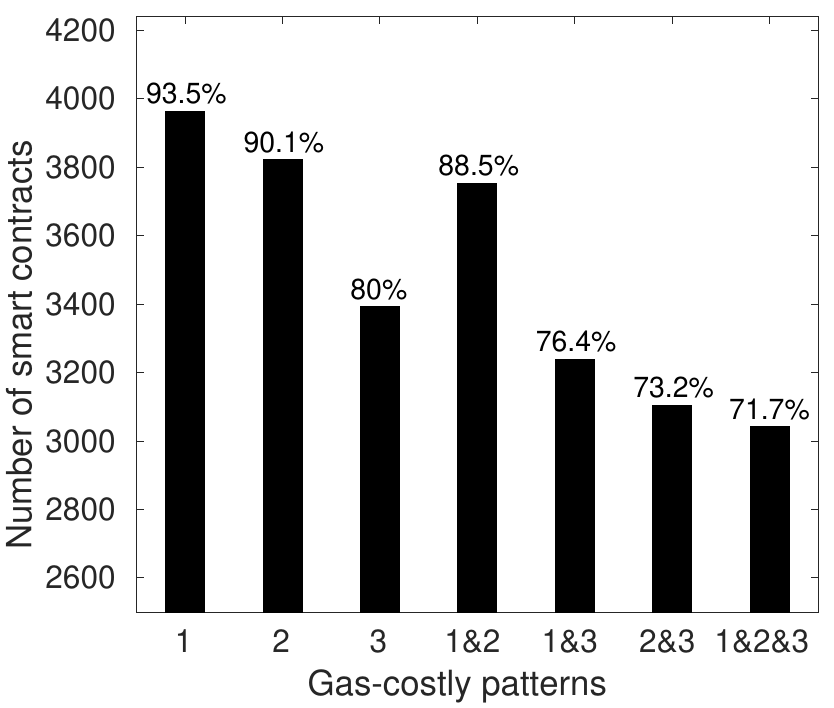}
	\vspace*{-1ex}
	\caption{Overview of gas-costly patterns: 1, 2, 3 indicate dead code, opaque predicates, and expensive operations in a loop, respectively.}
	\vspace{-2ex}
	\label{result}
\end{figure}
The number of smart contracts that have the 3 gas-costly patterns are illustrated in Fig.\ref{result}. More than 70\% contracts contain all these patterns, indicating that their bytecodes have not been properly optimized for reducing gas. Besides, more than 90\% contracts have  dead code or opaque predicates.



\begin{figure}[ht]
	\centering
	\vspace*{-3ex}
	\includegraphics[width=2.1in]{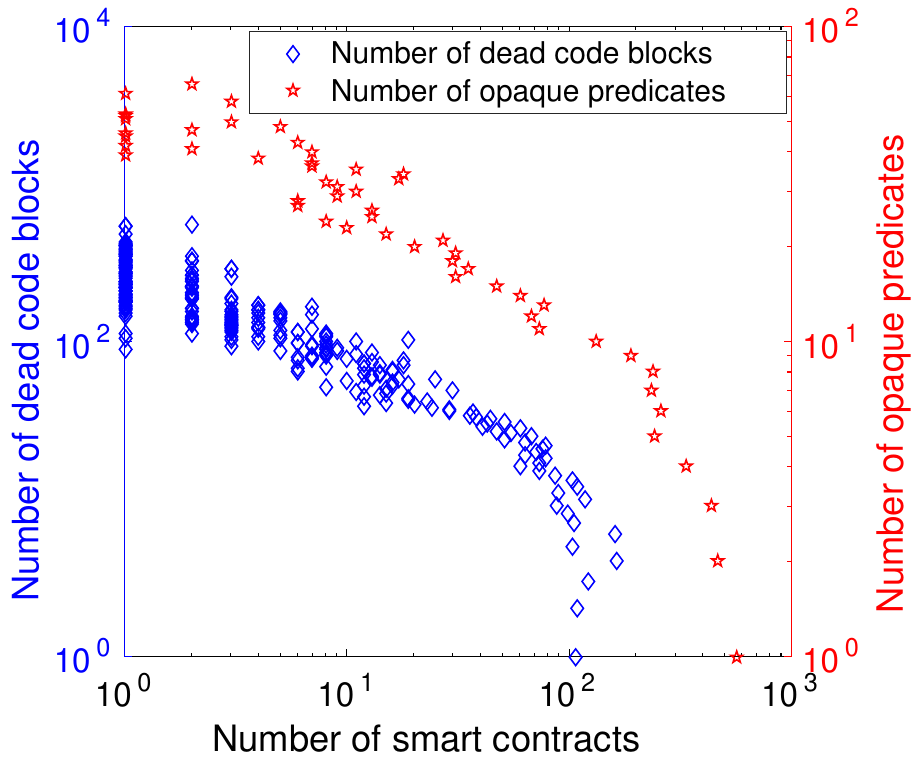}
	\vspace*{-1ex}
	\caption{Distribution of dead code blocks and opaque predicates in smart contracts.}
	\vspace{-2ex}
	\label{sta_dead}
\end{figure}

Fig.\ref{sta_dead} presents the distribution of dead code blocks and opaque predicates in smart contracts. Each point $(a,b)$ indicates that $a$ smart contracts contain $b$ dead code blocks or opaque predicates. Note that the contracts without these two patterns are not counted. The distributions of dead code blocks and  opaque predicates demonstrate similar trends: 51.7\% contracts contain more than 20 dead code blocks and 52.6\% contracts contain more than 10 opaque predicates. 

\begin{figure}[ht]
	\centering
	\vspace*{-3ex}
	\includegraphics[width=2.10in]{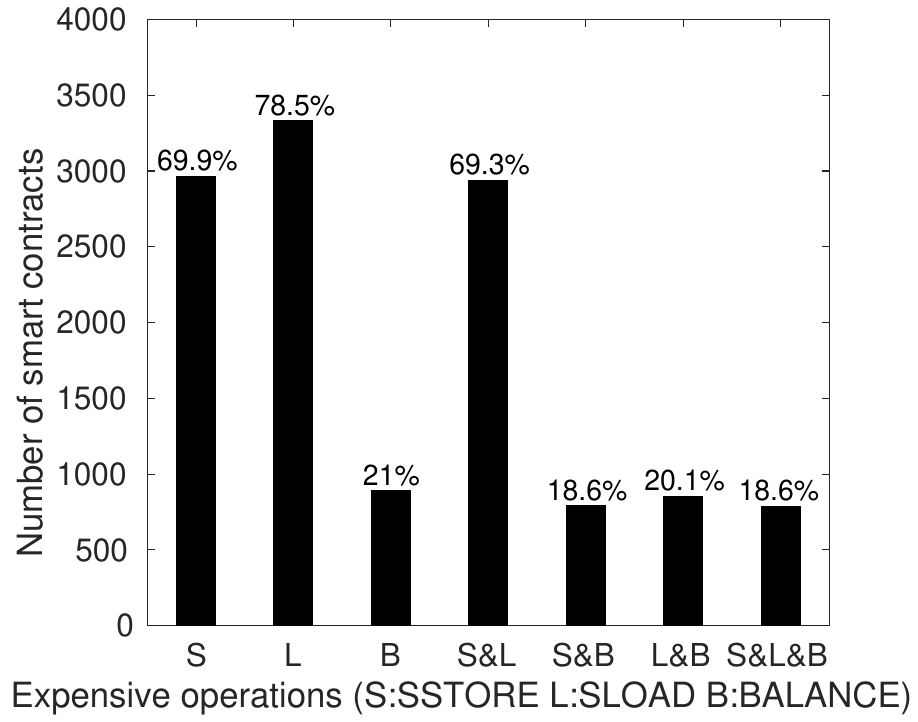}
	\vspace*{-1ex}
	\caption{Number of contracts containing expensive operations}
	\vspace{-5ex}
	\label{sta_exp}
\end{figure}

Fig.\ref{sta_exp} demonstrates that  69.9\%, 78.5\% and 21\% contracts have \texttt{SLOAD}, \texttt{SSTORE} and \texttt{BALANCE} operations in a loop, respectively. Moreover, if a contract has \texttt{SSTORE} operations in a loop (the percentage is 69.9\%), it may contain \texttt{SLOAD} operations (69.3\%) as well. Interestingly, if a contract uses \texttt{BALANCE} operations in a loop  (21\%), it likely contains both \texttt{SLOAD} and \texttt{SSTORE} operations (18.6\%).

\begin{figure}[ht]
	\centering
	\vspace*{-2ex}
	\includegraphics[width=2.30in]{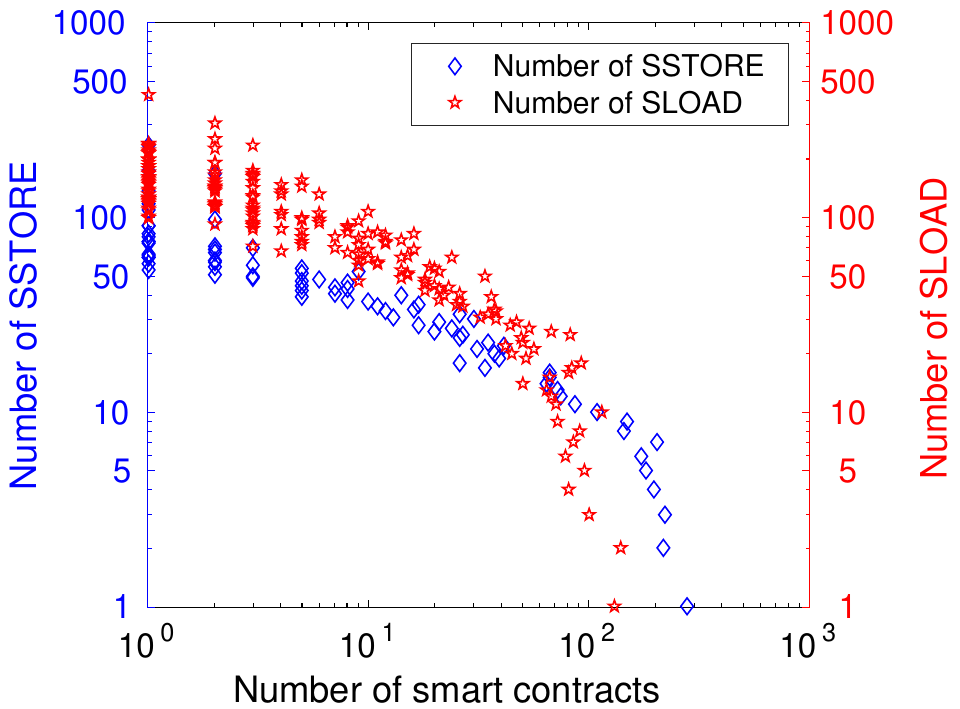}
	\vspace*{-1ex}
	\caption{Distribution of \texttt{SSTORE} and \texttt{SLOAD} within a loop in smart contracts.}
	\vspace{-3ex}
	\label{sta_sstore}
\end{figure}

Fig.\ref{sta_sstore} shows that a large number of contracts contain many expensive operations in a loop. For example,  57.1\% and 51.5\% of contracts have more than 7 \texttt{SSTORE} and 20 \texttt{SLOAD} operations in a loop, respectively. Note that contracts without such expensive operations in a loop are not counted. 

\begin{figure}[ht]
	\centering
	\vspace*{-2ex}
	\includegraphics[width=2.30in]{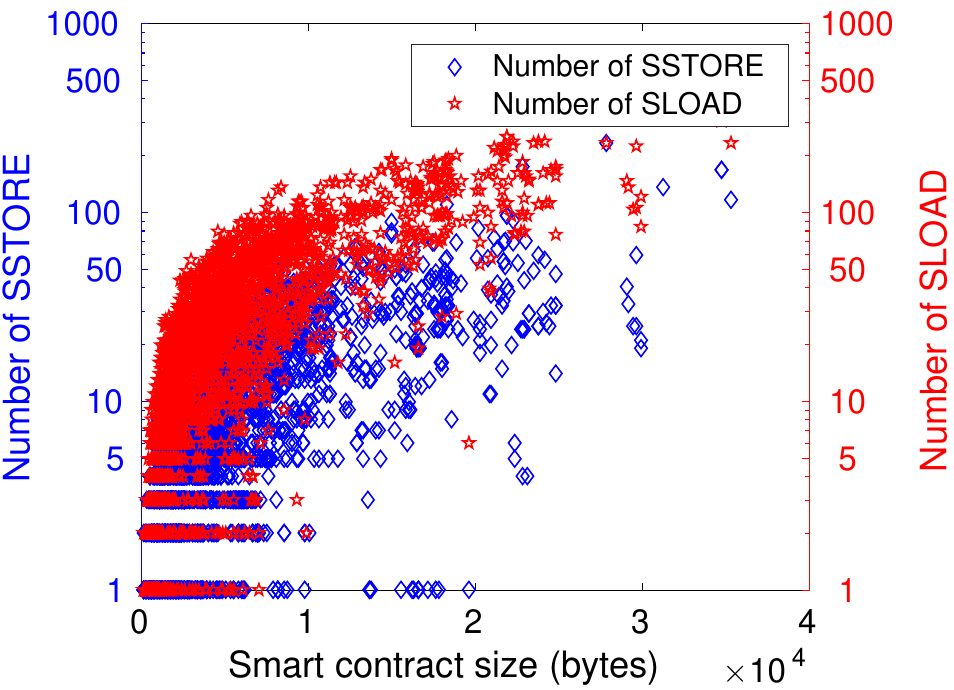}
	\vspace*{-1ex}
	\caption{Statistics of the size of contracts which contain \texttt{SSTORE} and \texttt{SLOAD} in a loop.}
	\vspace{-4ex}
	\label{sta_size}
\end{figure}

As expected, contracts with larger size are likely to contain more gas-costly patterns. Fig.\ref{sta_size} shows the relationship between the number of \texttt{SLOAD/SSOTRE} and the size of smart contracts. For example, a contract, named \emph{ARK}, which is of 34,767 bytes and deployed in 0x37b4869e73B7cE1284D6502B01aC81d500b50237, has 304 \texttt{SLOAD} and 168 \texttt{SSTORE} operations in loops.

\subsection{Real Case 1: FirstContract}
\begin{figure}[ht]
	\centering
	\vspace*{-1ex}
	\includegraphics[width=3.50in]{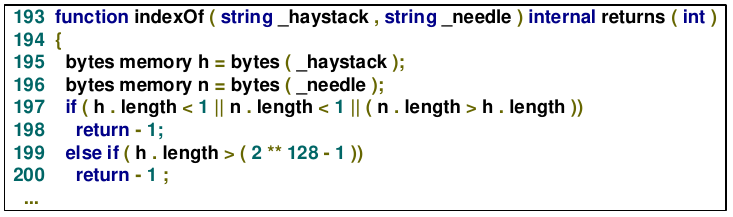}
	\vspace*{-3ex}
	\caption{Gas-costly code in \emph{FirstContract}}
	\vspace{-3ex}
	\label{FirstContract}
\end{figure}
\emph{FirstContract} is open source and deployed at the address \url{0x68C7147205A8bEB9D99fD19908b93462CdFfC60d}. \textsc{Gasper} discovers dead code at Line 200 (i.e., pattern 1) and an opaque predicate (i.e., pattern 2) at Line 199, as shown in Fig.\ref{FirstContract}. The function \emph{indexof} takes in two strings, \emph{\_haystack} and \emph{\_needle}. At Line 195, \emph{\_haystack} is converted into a set of bytes, \emph{h}. At Line 199, the length of \emph{h} is compared to $2**128-1$. However, the predicate will never be evaluated to true because ``$**$'' stands for exponential arithmetic. Consequently, the code at Line 200 cannot be executed.

\subsection{Real Case 2: Ballot}
\begin{figure}[ht]
	\centering
	\vspace*{-2ex}
	\includegraphics[width=3.30in]{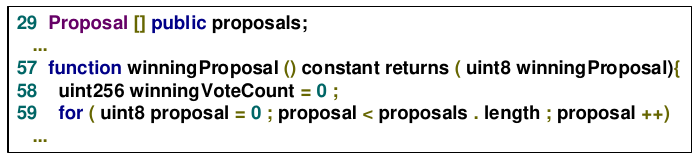}
	\vspace*{-1ex}
	\caption{Gas-costly code in \emph{Ballot}}
	\vspace{-2ex}
	\label{Ballot}
\end{figure}
\emph{Ballot} is also open source and deployed at the address \url{0x5A4964bb5FDd3CE646bB6AA020704F7D4db79302}. \textsc{Gasper} finds a \texttt{SLOAD} operation in a loop and it can be moved outside the loop, as shown in Fig.\ref{Ballot}.

Since the array \emph{proposals} (defined Line 29) is in the storage, getting access to its length (i.e., \textit{proposals.length} at Line 59) involves the  \texttt{SLOAD} operation. Moreover, the number of executing \texttt{SLOAD} is \textit{proposals.length}, because the length of \emph{proposals} is accessed in each iteration of the loop. This costly code can be optimized  by assigning \textit{proposals.length} to a stack variable, and then using the stack variable to do the comparison with \emph{proposal} at Line 59. After optimization, the number of using \texttt{SLOAD} can be reduced to only one.

\balance 
\section{Related Work}
\label{related}
There are a few studies on blockchain and smart contracts, but none of them investigates the gas consumption from the same viewpoint as ours. Luu et al. develop \textsc{Oyente}~\cite{oyente}, a novel symbolic execution based tool, to discover security bugs in Ethereum smart contracts. Bhargavan et al. use formal verification to analyze smart contracts (e.g., whether contracts check the return value of a \textit{send} operation because \textit{send} may fail)\cite{formal,why3}. \textsc{Hawk}~\cite{hawk} is a decentralized smart contract system enabling developers to write privacy-reserved smart contracts. Juels et al. find that smart contracts can facilitate crimes~\cite{ring} and show how criminal smart contracts can facilitate leakage of confidential information, theft of cryptographic keys, and various real world crimes. \textsc{Town Crier}~\cite{town} aims to provide trustworthy data to smart contracts because many applications of smart contracts need data from outside the blockchain. Atzei et al. survey a series of attacks which exploit the vulnerabilities of contracts to steal or tamper the assets~\cite{survey}.

\section{Conclusion And Future Works}
\label{conclusion}

We perform the \emph{first} investigation to expose that lots of smart contracts, generated by the recommended compiler Solidity, contain gas-costly bytecodes, which can be replaced with gas-efficient bytecodes to save money.  In particular, we identify 7 gas-costly patterns belonging to 2 categories.  Moreover, we propose and develop \textsc{Gasper} that leverages symbolic execution to automatically discover 3 representative gas-costly patterns in bytecode. By applying \textsc{Gasper}  to all deployed smart contracts until Nov. 5th, 2016, we find that 93.5\%, 90.1\% and 80\% smart contracts suffer from these 3 patterns, respectively. In future work, we will extend this research from the following aspects: (1) identifying more gas-costly patterns and the corresponding gas-efficient patterns; (2) extending \textsc{Gasper}  to cover all these patterns; (3) improving  compilers to produce gas-efficient bytecode. 



%

\section*{Acknowledgement}
This work is supported in part by the Hong Kong GRF (PolyU 152279/16E), the HKPolyU Research Grants (G-YBJX), Shenzhen City Science and Technology R\&D Fund (No. JCYJ20150630115257892), the National Natural Science Foundation of China (No.61402080, No.61572115, No.61502086, No.61572109), and China Postdoctoral Science Foundation founded project (No.2014M562307).





\bibliographystyle{IEEEtran}
\bibliography{ref.bib}
%


\end{document}